\documentstyle{article}
\newcommand{\ds}{\displaystyle}
\newcommand{\Vec}[1]{\vec{#1}}
\newcommand{\ghat}{\hat{g}}
\newcommand{\gprime}{g'}
\newcommand{\be}{\begin{equation}}
\newcommand{\ee}{\end{equation}}
\newcommand{\bea}{\begin{eqnarray}}
\newcommand{\eea}{\end{eqnarray}}
\newcommand{\abs}[1]{{\left\vert #1 \right\vert}}
\newcommand{\Nabla}{\vec\nabla}
\title{MOND, dark matter, and conservation of energy}
\author{Ahmad Shariati, Nosratollah Jafari
\\[2mm] \textit{Department of Physics, Alzahra University,}
\\[0mm] \textit{Tehran 19938-91167, Iran.}
\thanks{Emails: \texttt{shariati@mailaps.org, njafary@iasbs.ac.ir} }}
\date{30 Sep 2007}
\begin{document}

\newpage
\maketitle
\begin{abstract}
The MOND equation $m \, \vec a \, \mu(a) = \vec F$ could be
transformed to the equivalent form $m \, \vec a = \vec{F'}$,
where $\Vec{F'}$ is a transformed force.  Using this transformation
we argue that MOND could not avoid introducing dark matter, and
introduces nonconservative terms to the equations of motion.
\end{abstract}

\section{Introduction}
To solve the plateau of rotation curves of spiral galaxies there
are two paradigms, Dark Matter (DM) and Modified Newtonian
Dynamics (MOND).  MOND was proposed in 1983 by M.~Milgrom
\cite{Milgrom83}. The basic idea of modifying
Newtonian dynamics is to write the governing differential
equations not as the usual Newtonian form $\Vec{F} = m \,
\Vec{a}$, but as the equation $\Vec{F} = m \, \Vec{a} \, \mu(a)$,
where $\mu$ is a monotonically increasing function of $a = \abs{\vec{a}}$
such that for large enough values of
$a$ (compared to some fundamental acceleration of the theory
$a_0$), $\mu(a) \simeq 1$, and for very small values of $a$,
$\mu(a) \simeq 0$. Milgrom showed that this modification of the
Newtonian dynamics, for an $a_0 \sim 10^{-10}\, {\rm m}/{\rm s}^2$
could account for the flatness of the rotation curves of spiral
galaxies, with no need of introducing any extra (dark) matter.

To choose one of the alternative paradigms, DM or MOND, various
groups proposed or did experiments. Recently Gundlach and
colleagues \cite{GSSC07} have shown for accelerations as small as
$10^{-14}\, {\rm m}/{\rm s}^2$ the Newtonian equation $\Vec{F} =
m\, \Vec{a}$ is valid.  (Two decades ago A.~Abramovici, Z.~Vager
\cite{AV86} showed that for the accelerations as small as
$10^{-11}\, {\rm m}/{\rm s}^2$ the Newtonian equation is valid.)

Apart from these experiments, there are theoretical reasons to
abandon MOND paradigm.  One is that MOND is somehow equivalent to
DM (see for example Dunkel's \cite{Dunkel04}). Here, we
would like to add to this theoretical argument.

\section{Transforming MOND differential equations to Newtonian ones}
The basic idea behind this article, is that $\Vec{F} = m\, \Vec{a}$ is a framework to write
the dynamics (see, for example Wilczek's \cite{Wilczek04a}).
The most important implication of $\Vec{F} = m\, \Vec{a}$ is that the differential equation governing a
point particle is of second order, such that when put in the form $\Vec{F} = m\, \Vec{a}$, the \textit{force}
$\vec F$ depends on position and velocity of the particle, and not on the acceleration $a$ itself.
Now, MOND differential equation is $\Vec{F} = m \, \Vec{a} \, \mu(a)$.  Why not writing it in the usual
Newtonian form $\Vec{F}' = m\, \vec a$, for some other force $\Vec{F}'$?  Let's do that.

Let $\mu(x)$ be a smooth, positive, and monotonically increasing
function defined for $x \geq 0$, and such that $\ds{\lim_{x\to 0}
\mu = 0}$, and $\ds{\lim_{x\to\infty} \mu = 1}$, and $\mu'(x) > 0$
(here $\mu'(x)$ is the derivative of $\mu$ with respect to its
argument $x$).  Then, one can always solve the equation $x \,
\mu(x) = y$ for $x$, getting $x = y \, \eta(y)$, where $\eta$ is a
smooth function of $y$. The proof is a simple application of the
well known inverse function theorem. All we have to show is
that the derivative of $x\, \mu(x)$ never vanishes for $x > 0$,
and that's trivial.

Dividing $x \, \mu(x) = y$ by $x = y \, \eta(y)$, one get $\mu(x)
\, \eta(y) = 1$, from which it follows that $\eta(y)$ is a
monotonically decreasing function, asymptotic to the constant 1,
for large $y$.

Now consider MOND vector equation of motion $\Vec{a} \, \mu(a) =
\Vec{g}$, where $\Vec{a}$ is the acceleration, $a = \abs{\Vec{a}}$,
and $\Vec{g} := \Vec{F}/m$ is a function of velocity $\Vec{v}$,
position $\Vec{r}$, and mass $m$ of the particle. Squaring this
equation and then taking the square root (noting that everything
is positive), we get $a \, \mu(a) = g$.  Solving this equation for
$a$, we get $a = g \, \eta(g)$. Now denote the direction of
$\Vec{g}$ by $\ghat$, that is $\Vec{g} = g\, \ghat$, and  note
that $\Vec{a} = a \, \ghat$.  Multiply $a = g \, \eta(g)$ by
$\ghat$ to obtain $\Vec{a} = \Vec{g} \, \eta(g)$ This is the usual
Newton's equation of motion, for a modified acceleration field
$\Vec{g} \, \eta(g)$

Example 1. The \textit{simple} form of the MOND function $\mu$:
\be \mu(a) = \frac{a}{a + a_0} \ee
\be a \, \mu(a) = \frac{a^2}{a + a_0} = g \hskip 5mm \Rightarrow \hskip 5mm
a = \frac g 2 \left( 1 + \sqrt{1 + \frac{4\, a_0}{g}} \, \right). \ee
\be \Vec{a} = \Vec{g}\, \eta(g) \hskip 10mm
\eta(g) := \frac 1 2 \, \left( 1 + \sqrt{1 + \frac{4\, a_0}{g}} \, \right) \ee

Exapmple 2. The \textit{standard} form of the MOND function $\mu$:
\be \mu(a) = \frac{a}{\sqrt{a^2 + a_0^2}}~~\Rightarrow~~ a \,
\mu(a) = \frac{a^2}{\sqrt{a^2 + a_0^2}} = g  \ee \be \Vec{a} =
\Vec{g} \, \eta(g) \hskip 10mm \eta(g) := \frac{1}{\sqrt{2}}
\left( 1 + \sqrt{1 + \frac{4 \, a_0^2}{g^2}}\, \right)^{1/2} \ee

Up to this point, we have shown that the MOND equation $\Vec{a} \,
\mu(a) = \Vec{g}$ could be written as $\Vec{a} = \Vec{g} \,
\eta(g)$, which is the usual Newtonian equation of motion, but for
the modified acceleration (or force) field $\Vec{\gprime} := \Vec{g} \,
\eta(g)$. Because we have quite a lot of experience with this
Newtonian equation, and because it is completely equivalent to the
MOND equation $\Vec{a} \, \mu(a) = \Vec{g}$, we could now derive
some useful information about MOND theories.

Consider the MOND equation for a gravitational field $\Vec{g}$,
where $\Vec{g}$ solves the usual Newtonian field equations
$$\Nabla\cdot\Vec{g} = - 4 \, \pi \, G \, \rho_m, \qquad
\Nabla\times\Vec{g} = 0~,$$
 where $\rho_m$ is the \textit{mass density}
function.  Define $\Vec{\gprime} = \Vec{g} \, \eta(g)$.  The MOND
equation is equivalent with $\Vec{a} = \Vec{\gprime}$.   Now let's
investigate the differential equations for $\Vec{\gprime}$. 
\bea
\Nabla\cdot\Vec{\gprime} & = & - 4 \, \pi \, G \, \rho_m \, \eta(g) +
\Vec{g} \cdot \Nabla\eta(g),
\\ \Nabla\times\Vec{\gprime} & = & - \Vec{g} \times \Nabla \eta(g).
\eea

We see that the mass density is being modified (multiplied by
$\eta(g)$), and we have got an extra term in
$\Nabla\cdot\Vec{\gprime}$, which could be interpreted as a
\textit{dark} mass density. As an example, let's consider the
acceleration round a point mass $M$, where $\Vec{g} = -
\ds{\frac{G\, M}{r^2}} \hat r$ for which $g = G M /r^2$. For
simplicity, let's take the simple form of the MOND function $\mu$.
By a straightforward calculation we get 
\bea \frac{-1}{4\, \pi \, G} \Nabla\cdot\Vec{\gprime} 
= M \, \delta(\Vec{r}) + \frac{a_0}{2\, \pi\, G} \,
\frac{1}{r} \, \left(1 + \frac{4\, a_0\, r^2}{G\, M}
\right)^{-1/2} \eea This clearly shows that accepting MOND
equation $\Vec{a} \, \mu(a) = \Vec{g}$ for Solar System (or any
star), is equivalent to accepting an infinite dark matter, with
density $\rho_d= a_0/\left(2\pi G\, r \, \sqrt{1 + 4\, a_0 \, r^2/G\,
M}\right)$.

If $\rho_m(\Vec{r})$ has spherical symmetry, then $\Vec{g}$ is
always in the direction of $-\hat r$, therefore $\Vec{g} \times
\Nabla \eta(g)$ vanishes.  But in general, we do not have
spherical symmetry (for example for an elliptical galaxy) so that
$\Vec{g} \times \Nabla \eta(g)$ does not vanish. This means that
$\Nabla\times\Vec{\gprime}\neq 0$.  This would imply profound
effects, because from $\Vec{a} = \Vec{\gprime}$, one could easily
get the Work-Kinetic Energy theorem \be \Delta \left( \frac 1 2 \,
m \, v^2 \right) = m \, \int \Vec{\gprime} \cdot d\Vec{\ell}. \ee
If $\Nabla\times\Vec{\gprime} \neq 0$, then the line integral
depends on path, and in particular it does not vanish for a closed
path.  This suggests the following method to find observable
results.

Let's write MOND equation $m \, \vec a \, \mu(a) = m\, \vec g$ in
the equivalent form $m\, \vec a = m\, \Vec{\gprime}$, and write
$\Vec{\gprime} = \vec g + \vec{\delta g}$.  It is easily seen that
now we have \be \Delta \left( \frac 1 2 \, m \, v^2 + V \right) =
m \, \int \Vec{\delta g} \cdot d\Vec{\ell}, \ee where $V$ is the
usual Newtonian potential energy ($\Vec g = - \Nabla V$). If
$\vec{\delta g}$ is small, we can consider the right hand side as
a perturbation and deduce some observable results. To get a
feeling, consider the circular orbit of a satellite round a
planet.  In the Newtonian mechanics $\vec \delta g$ vanishes, and
$E := \frac 1 2 \, m \, v^2 + V$ is conserved, which, for a
circular orbit means that both $v$ and $r$ are constants.  Now, a
small, but non-vanishing $\Vec{\delta g}$, could lead to a change
in $E$, and this results a change in $r$.  For small enough
$\Delta E$, one can apply Newtonian results as follows.

For a circular orbit we have $\frac{m\, v^2}{r} = \frac{G\, M \, m}{r^2}$,
from which it follows that $E = - \frac{G\, M\, m}{2\, r}$, and therefore,
\be \label{dedr} \frac{\Delta r}{r} = - 2 \frac{\Delta E}{E}. \ee
MOND would imply that there is a deviation from Newtonian mechanics, such that
\be \Delta E := m \, \oint \Vec{\delta g} \cdot d\Vec{\ell}. \ee
This may lead to some observable results.

\section{Conclusion}
A point of view accepted by many physicist is that $\Vec F = m\, \Vec a$ is
a framework to write equations of dynamics.  Accepting this point of view, 
the MOND equations $\Vec F = m \, \Vec a \, \mu(a)$ could be
transformed to $\Vec F' = m \, \Vec a$, where $\Vec F'$ is a
transformed force vector.  From this form of the equation of
motion it follows that MOND could not avoid introducing dark matter.
Besides, it violates conservation of energy. Here, by energy
we mean $\frac 1 2 \, m \, v^2 + V$, where $V$ is the usual Newtonian
potential energy.

\section*{Acknowledgements}  We would like to thank A.-H. Fatollahi for
his valuable comments.

\end{document}